\RequirePackage{fix-cm}
\documentclass[smallextended]{svjour3}
\smartqed
\usepackage{amsmath,amssymb,paralist,subfigure,graphicx,cite}
\usepackage{algorithm2e}
\usepackage{color}

\usepackage{url}            % simple URL typesetting
\usepackage{enumerate}   
\usepackage{pifont}
\usepackage[usenames,dvipsnames]{xcolor}
\usepackage[utf8]{inputenc} % allow utf-8 input
\usepackage{booktabs}       % professional-quality tables
\usepackage{amsfonts}       % blackboard math
\usepackage{nicefrac}       % compact symbols for 1/2, etc.
\usepackage{microtype}      % microtypography
\usepackage{lipsum}
\usepackage{graphicx}
\usepackage{dsfont}
\usepackage{enumitem}

\def\mc{\mathcal}

\begin{document}
	
	\title{Impact of Clustering on the Observability and Controllability of Complex Networks
	}
	\author{Mohammadreza Doostmohammadian$^\ast$  \and
		Hamid R. Rabiee
	}
	
	\institute{ M.  Doostmohammadian \at
		Department of Mechatronics, Faculty of Mechanical Engineering, Semnan University, Semnan, Iran ($^\ast$ corresponding author). \\
		Tel.: +98-23-31533429\\
		\email{doost@semnan.ac.ir}		
		\and		
		H. R. Rabiee \at
		Department of Computer Engineering, Sharif University of Technology, Tehran, Iran. \\ 
		\email{rabiee@sharif.edu}
		}
	
	\date{Received: date / Accepted: date}

	\maketitle
	
	\begin{abstract}
	The increasing complexity and interconnectedness of systems across various fields have led to a growing interest in studying complex networks, particularly Scale-Free (SF) networks, which best model real-world systems. This paper investigates the influence of clustering on the observability and controllability of complex SF networks, framing these characteristics in the context of structured systems theory. In this paper, we show that densely clustered networks require fewer driver and observer nodes due to better information propagation within clusters. This relationship is of interest for optimizing network design in applications such as social networks and intelligent transportation systems. We first quantify the network observability/controllability requirements, and then, through Monte-Carlo simulations and different case studies, we show how clustering affects these metrics. Our findings offer practical insights into reducing control and observer nodes for sensor/actuator placement, particularly in resource-constrained setups. This work contributes to the understanding of network observability/controllability and presents techniques for improving these features through alterations in network structure and clustering.
		
	\end{abstract}

\section{Introduction}	
The study of complex networks has gained significant attention across various fields due to its ability to model interactions (as links) within systems comprising numerous interconnected elements (as nodes). This ranges from IoT networks \cite{lcss_cluster} to social networks \cite{urena2019social} and transportation systems \cite{wang2016complex,pirani2022impact}. In particular, Scale-Free (SF) networks are known to best model these real-world systems. Among the key characteristics of these networks are observability \cite{liu2013observability} and controllability \cite{liu2011controllability}, which refer to the capacity to infer the states of the system from measured outputs and to influence the system behaviors via input signals, respectively. Many works have studied different structural and functional properties of complex networks \cite{boccaletti2006complex}.  Among these, clustering--often seen as densely connected groups within the broader network topology--emerges as a key feature that influences different spreading processes and optimization behaviours of the underlying system, see \cite{snam,snam_opt} for example. This feature, for example, defines the vulnerability of the network to external perturbations such as random node failures \cite{kuhnle2017vulnerability} and small-worldness of industrial networks \cite{prabhakar2021exploration}.  

In this paper, we study how clustering affects the observability/controllability of complex SF networks. These networks may model complex systems, where observability/controllability analysis is based on structured systems theory by modelling these systems as adjacency graphs \cite{liu2016control,liu2011controllability}. In such networks, clustering can significantly affect how information propagates across a network and how effectively it can be controlled/observed. Particularly in social networks, where individuals or entities interact via complex dynamics, the clustering of users can affect the spread of information, making it critical for applications such as targeted marketing, viral phenomena, and opinion dynamics. Observability analysis can help researchers analyze community influence and key inference nodes, while control strategies can guide the network towards the desired state, such as consensus, agreement, or other collaborative efforts \cite{urena2019social,mameli2022social}. Similarly, in intelligent transportation networks clustering manifests in the configuration of vehicles and infrastructure, impacting traffic flow and safety. As vehicles communicate and coordinate through local interactions, understanding the clustering effects can enable better traffic management, accident prevention, and resource allocation \cite{pirani2017graph,badnava2021platoon}. The study of observability and controllability in these settings is vital for driving these systems toward the desired state \cite{isj19minimal} or estimating their states via minimal sensor placement \cite{spl18}.

In this work, we quantify the ability to control/observe a complex network as its minimum number of required driver/observer nodes. We particularly show that densely clustered networks require less number of driver/observer nodes. This is because as nodes are densely interconnected, information/input-signals can spread quickly within clusters, which can facilitate achieving global observability and controllability. Conversely, sparsely connected networks, which are less interconnected within clusters, require more driver/observer nodes for control and observation. Understanding these is essential for the scheduling of large-scale real-world networks in different applications.
By illustrating observability/controllability requirements on large-scale networks, this paper provides insights into optimizing network design and functionality in diverse applications. In this direction, one can use the existing algorithms for tuning the clustering coefficient \cite{liu2018optimization,kashyap2018mechanisms} to reduce the number of driver/observer nodes in complex networks and enhance their controllability/observability via link addition. In this direction, we provide a technique to change the structure of the complex networks and their clustering to reduce the number of control/observer nodes. Note that control and observation nodes often require resources such as energy, processing power, and maintenance \cite{zografopoulos2021cyber}. In many real-world scenarios, especially in resource-constrained environments (e.g., sensor networks, IoT applications, transportation systems), minimizing the number of such nodes can lead to significant cost savings.    

The rest of the paper is organized as follows. Section~\ref{sec_prob} states the main problem and provides a metric for the observability/controllability of networks. Section~\ref{sec_net} defines the main properties of SF networks. Section~\ref{sec_mc} presents the main empirical results based on Monte-Carlo simulation to study the clustering feature of SF networks. Section~\ref{sec_case} provides some real-world networks as case studies to explore observability/controllability versus clustering. Finally, Section~\ref{sec_con} concludes the paper.

\section{The Observability/Controllability Problem Statement} \label{sec_prob}
%In this section, first the notions of structural observability/controllability over graphs are described, the related metric is defined, and the problem is stated based on these notions.
Observability in networks refers to the ability to infer the state of the entire network based on some subset of its nodes. Controllability, on the other hand, is the capability to drive the state of the system from any initial state to any desired final state through suitable control inputs. Both concepts are important in understanding the dynamics and functionality of networks, especially in systems such as social networks, transportation networks, communication systems, and dynamic systems. The concepts of controllability and observability of networks are analyzed based on graph-theoretical principles as discussed in \cite{liu2011controllability, liu2013observability, tnse18, isj19minimal}, which build upon the structural methodology presented in \cite{lin}. In this section, we describe relevant graph-theoretic concepts to provide a metric quantifying the observability/controllability of networks and then formulate the problem in this paper.

\begin{figure}[hbpt!]
	\centering
	\includegraphics[width=3.5in]{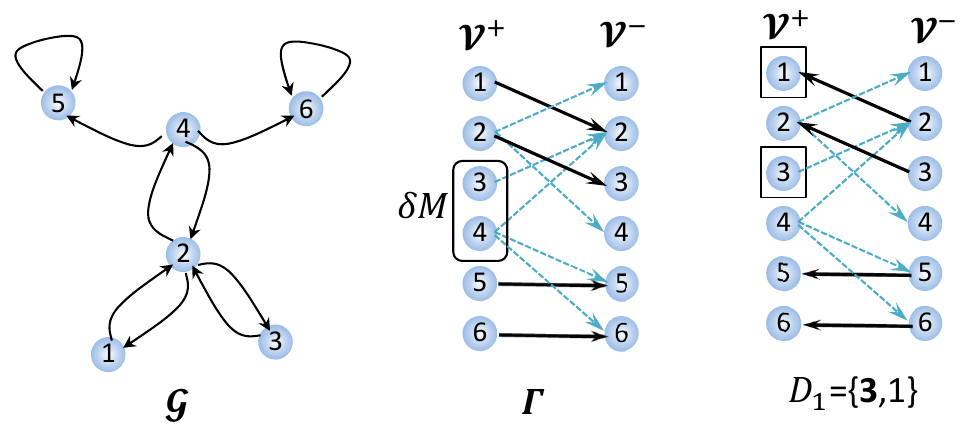}
	\caption{This figure shows the graph $\mc{G}$ and its bipartite representation $\Gamma$. In the bipartite graph $\Gamma$, the black links represent one possible matching, with $\delta\mc{M}$ as its set of unmatched nodes. An example dilation $\mc{D}_1$ found via Dulmage-Mendelsohn decomposition is presented in the right figure.}
	\label{fig_matching}
\end{figure}

A network is depicted as a graph $\mc{G}=(\mc{V},\mc{E})$ where $\mc{V}=\{1,...,n\}$ denotes the set of nodes and $\mc{E}=\{(i,j)|i \leftrightarrow j\}$ represents the set of links. It is important to note that the links are treated as bidirectional since the networks considered in this paper are undirected. The degree of node $i$ is represented as $d_i$ 
 and its neighboring nodes by $\mc{N}_i$. For structural analysis, we define a bipartite representation of the graph $\Gamma = (\mc{V}^+,\mc{V}^-,\mc{E}_\Gamma)$, with two sets of disjoint nodes as 
$\mc{V}^+=\mc{V}^-=\mc{V}$ and the set of links 
$\mc{E}_\Gamma=\{(i^+,j^-)|(i,j) \in \mc{E}\}$
all directed from $\mc{V}^+$ set to $\mc{V}^-$ set. This representation simplifies the analysis of network properties like matching. We define a max matching, denoted as $\mc{M}$, as the largest collection of links without any shared end-nodes and begin-nodes. In the bipartite representation 
$\Gamma$, the max matching corresponds to the maximum links connecting 
$\mc{V}^+$ to $\mc{V}^-$ without sharing starting nodes in 
$\mc{V}^+$
or ending nodes in 
$\mc{V}^-$. One may find different max matchings in the graph and it is not unique in general. For an in-depth understanding of these graph-theoretic concepts, we refer interested readers to \cite{murota}. Given this background, we describe the \textit{matched/unmatched nodes} which define the main entity in the graph related to its observability/controllability. The matched nodes, denoted as 
$\rho\mc{M}$, are defined as the nodes incident to 
$\mc{M}$, and the unmatched nodes are specified as  
$\delta\mc{M} = \mc{V}\backslash\rho\mc{M}$. Recall that the collection of unmatched nodes is not necessarily unique due to the inherent non-uniqueness of the matching component. The set of all equivalent unmatched nodes defines the contraction (or dilation) in the context of observability (and controllability)\footnote{The notions of controllability and observability are dual concepts in graph perspective, i.e.,  by reversing the links of a controllable graph the resultant graph is observable. Therefore, in the context of undirected graphs with bidirectional links controllability and observability are equivalent.}. We provided Fig.~\ref{fig_matching} to better illustrate these concepts. 
To find the max matching $\mc{M}$ and unmatched nodes $\delta\mc{M}$ in a graph, an efficient algorithm is given by \cite{galil1986efficient}, which is of order $\mc{O}(\log^3(n))$ complexity. One can find the contractions and dilations in the graph by using the \textit{Dulmage-Mendelsohn decomposition} algorithm \cite{dulmage1958coverings}.

To provide a metric for controllability/observability analysis, we quantify these terms by the number of driver nodes and observer nodes. Define a driver node as the target node within the network to inject the control input \cite{liu2011controllability, isj19minimal}. In a similar way, define an observer node as the target node to be inferred for the purpose of estimation \cite{liu2013observability,spl18}. Given that the network under consideration is undirected, the dual concepts of controllability and observability are interchangeable, leading to an equal count of observer nodes and driver nodes. It has been demonstrated that in the context of network controllability unmatched nodes function as driver nodes \cite{isj19minimal} and in the context of observability unmatched nodes function as observer nodes \cite{tnse18}. This implies that by applying control inputs to the unmatched nodes, the network can be controlled, and, equivalently, by observing the states of the unmatched nodes, the overall state of the network becomes observable to the estimator. Furthermore, the count of unmatched nodes defines the minimal number of inputs and outputs needed for controllability and observability, respectively. Thus, the number of unmatched nodes serves as a metric for assessing the controllability and observability of complex networks \cite{liu2012controlcentrality}.

Similar study in \cite{yuan2013exact} addresses the exact controllability of networks using the Popov-Belevitch-Hautus (PBH) rank test (equivalent to the Kalman rank condition) and shows that the minimum number of independent driver nodes equals the maximum geometric multiplicity across all eigenvalues of the adjacency (or network) matrix. For many large, sparse networks typical of real applications, this criterion reduces to a simple rank deficiency of the network matrix $A$, so that the required number of driver nodes can be expressed as $n - \mbox{rank}(A)$. On the other hand, the structural approach in \cite{liu2011controllability} (grounded in the theory surveyed in \cite{dion2003generic}) counts driver nodes as the number of unmatched nodes in a maximum matching of the network, which equals the structural (or generic) rank deficiency of $A$; equivalently the number of driver (or observer) nodes is $n - \mbox{Grank}(A)$, where ``Grank'' denotes the generic/structural rank defined in \cite{dion2003generic}. Importantly, \cite{dion2003generic} notes that for connected networks the generic rank equals the numerical rank for \textit{almost all} choices of the nonzero parameters (i.e., exceptions form an algebraic variety of Lebesgue zero measure), so the structural result in \cite{liu2011controllability} is consistent with the PBH-based conclusion of \cite{yuan2013exact} for typical sparse connected networks. Note that the controllability analysis in this paper follows the structural theory in \cite{liu2011controllability,dion2003generic} and holds for ``sparse'' networks.

To study the effect of clustering on observability and controllability, we count the number of observers and driver nodes (i.e., the number of unmatched nodes $|\delta\mc{M}|$) in complex networks. Two main types of SF networks are considered in this work: (i) the model proposed by Barab{\'a}si-Albert \cite{barabasi_albert1999} (referred to as BA model), and (ii) the model proposed by Holme-Kim \cite{Holme2002csf} (referred to as HK model). The main difference between these two models is in their clustering properties, while other characteristics such as power-law degree distribution and average shortest-path length are similar. As discussed in detail in Section~\ref{sec_net}, the HK model is more clustered based on its specific design compared to the BA model. Therefore, to study the effect of clustering on the controllability/observability of the network, we compare the number of unmatched nodes in BA and HK networks of the same size. Note that we keep other network properties such as average node degree and number of links the same for both models, and thus, the controllability/observability is only affected by the network clustering. The results of this work allow to manage the controllability/observability of large-scale networks (e.g., social networks or intelligent transportation networks) by tuning its clustering coefficient. To do this, we present a new structural technique that adds extra links to reduce the unmatched nodes and we examine this method on real-world case studies. This is significant as one can reduce the number of sensor/actuator nodes in real-world complex systems and reduce the associated costs of sensor/actuator placement. 

\section{Network Models with Tunable Clustering} \label{sec_net}
SF networks are characterized by a power-law degree distribution, meaning that a small number of nodes have a very high degree (many connections), while the majority have a low degree. This characteristic is often observed in real-world networks such as social and biological systems. Two prominent models for generating SF networks are the Barabási-Albert (BA) model \cite{barabasi_albert1999} and the Holme-Kim (HK) model \cite{Holme2002csf}. Below is a description of each model along with the algorithms used to construct these networks. The main difference between the two models is in their clustering coefficient which is described in Section~\ref{sec_clust}. 

\subsection{Barab{\'a}si-Albert Model}
The BA model \cite{barabasi_albert1999} introduces two key principles for the formation of SF networks: growth and preferential attachment. In this model, networks grow over time by adding new nodes. Each new node is more likely to connect to existing nodes that already have a high degree. This \textit{rich-get-richer} phenomenon leads to the emergence of hubs--nodes with many connections. The algorithm is composed of three steps as follows:

\begin{enumerate}
	\item \textbf{Initialization:} Start with an initial connected network of $m_0$ nodes as the small seed graph. Each node has a certain number of links (connections). 
	\item \textbf{Growth:} At each time step, add a new node to the network:
	The new node connects to $m$ existing nodes (where $1\leq m \leq m_0$) with a probability proportional to the degree of the nodes. This is called preferential attachment and means that nodes with higher degrees are more likely to be selected for connection\footnote{This preferential attachment in the growth procedure follows the power-law degree distribution of the SF networks, implying few number of hubs (high-degree nodes) and many low-degree nodes. This plays a key role in the algorithmic design in this paper, where the results specifically hold for SF networks.}.
	\item \textbf{Repeat:} Continue adding nodes until the desired network size 
	$n$ is reached.	
\end{enumerate}

The BA model results in a network with a power-law degree distribution, where the probability $P(d)$ of a node having $d$ links 
follows   $P(d) = d^{-\alpha}$ with $2<\alpha <3$ \cite{barabasi2003scale}.

\subsection{Holme-Kim Model}
The HK model \cite{Holme2002csf} was introduced as an extension of the BA model. It incorporates an additional mechanism called \textit{triad formation}, allowing for the creation of triangles between nodes. This modification captures real-life networks better by accounting for the tendency of nodes to cluster together, as observed in social networks \cite{Toivonen2006social}. The following steps explain the building blocks of the HK model:

\begin{enumerate}
	\item \textbf{Initialization:} Similar to the BA model, begin with a small connected network of $m_0$ nodes. 
	\item \textbf{Growth:} At each time step, add a new node to the network:
	The new node first connects to $(1-p)m$ existing nodes with a probability proportional to their degrees, following the preferential attachment mechanism (as in the BA model). The parameter $p$ is defined next.
	\item \textbf{Triangle Formation:} After the new node connects to its $(1-p)m$ chosen nodes, it has an additional chance to create links to nodes that are neighbors of the nodes it just connected to. This is typically determined by parameter $p$ which defines the fraction of the newly created links that will be made to neighbors of the already-connected nodes. In this step, $pm$ new links are added to the network, and overall $m$ links are added by every new node to the old nodes in both growth and triangle formation steps. 
	\item \textbf{Repeat:} Continue adding nodes until reaching the desired network size $n$.	
\end{enumerate}

The HK model also yields a power-law degree distribution, but it produces a network with clustering coefficients that better reflect real-world phenomena, hence offering a more realistic portrayal of social networks. Note that, in contrast to the BA model emphasizing the importance of preferential attachment and network growth, the HK model includes features that account for clustering \cite{sidorov2021growth}.

\subsection{Clustering Coefficient} \label{sec_clust}
The clustering coefficient is a measure used to quantify the degree to which nodes in a graph tend to cluster together \cite{kaminski2021mining}. It provides insight into the local structure of the network by indicating how interconnected the neighbors of a node are. A common way to define the global clustering coefficient is by considering the ratio of the number of closed triplets (or triangles) to the number of total triplets in the graph.
As shown in \cite{szabo2003structural}, in the BA model the clustering coefficient is independent of the node degree $d$ and is calculated by the following:
\begin{equation}\label{eq_C_BA}
	C(d) = \frac{m-1}{8}\frac{(\log(n))^2}{n}.
\end{equation}	
with $n$ as the network size and $m$ as the number of nodes which get connected to the newly added node in the growth step for BA model.
For the HK model, the clustering coefficient follows as \cite{szabo2003structural}:
\begin{equation} \label{eq_C_HK}
	C(d) \approx  \frac{4m_1}{d}+\frac{m-1}{8}\frac{(\log(n))^2}{n},
\end{equation}
with $m_1 := pm$ as the number of connected nodes in the triangle formation step for HK model and $d$ as the node degree. It is clear from these equations that the clustering coefficient of the HK model is greater than the BA model of the same size $n$ and the same average node degree $2m$, which is due to the term $ \frac{4m_1}{d}$ following the triad formation  \cite{Holme2002csf,szabo2003structural}.
	 
\section{Empirical Results based on Monte-Carlo Simulation} \label{sec_mc}
In this section, we perform Monte-Carlo simulation on BA and HK networks of different sizes to compare their observability/controllability metric. Recall that both types of networks follow power-law degree distribution and logarithmically increasing average shortest path while they have different clustering properties. In the simulation, we also keep the number of links (in the network of the same size) and average node degree the same in both types to solely study the effect of clustering coefficient.

\begin{figure}[hbpt!]
	\centering
	\includegraphics[width=3in]{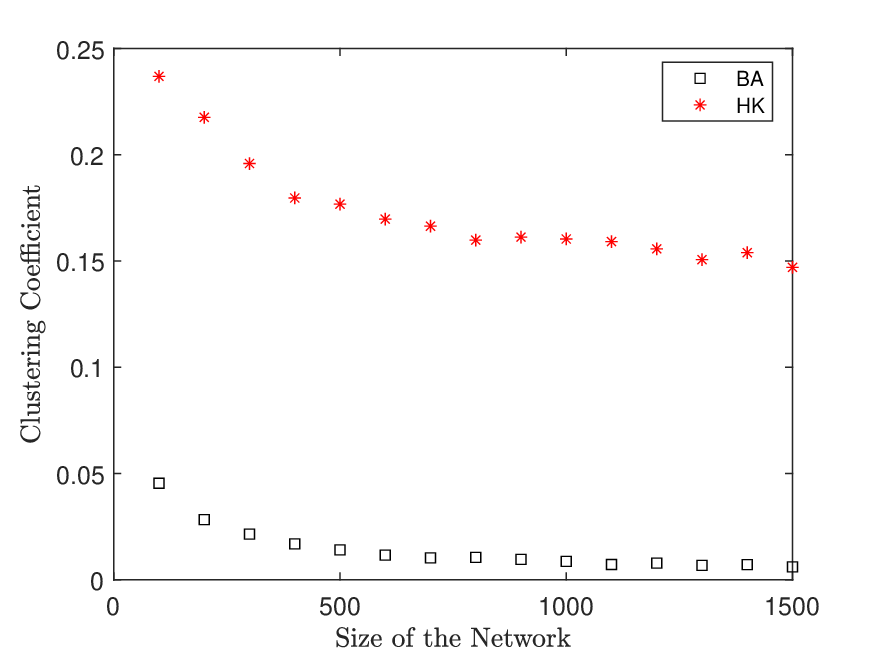}
	\caption{This figure compares the clustering coefficient of the BA and HK networks of the same size.}
	\label{fig_clust_size}
\end{figure}

\begin{figure}[hbpt!]
	\centering
	\includegraphics[width=3in]{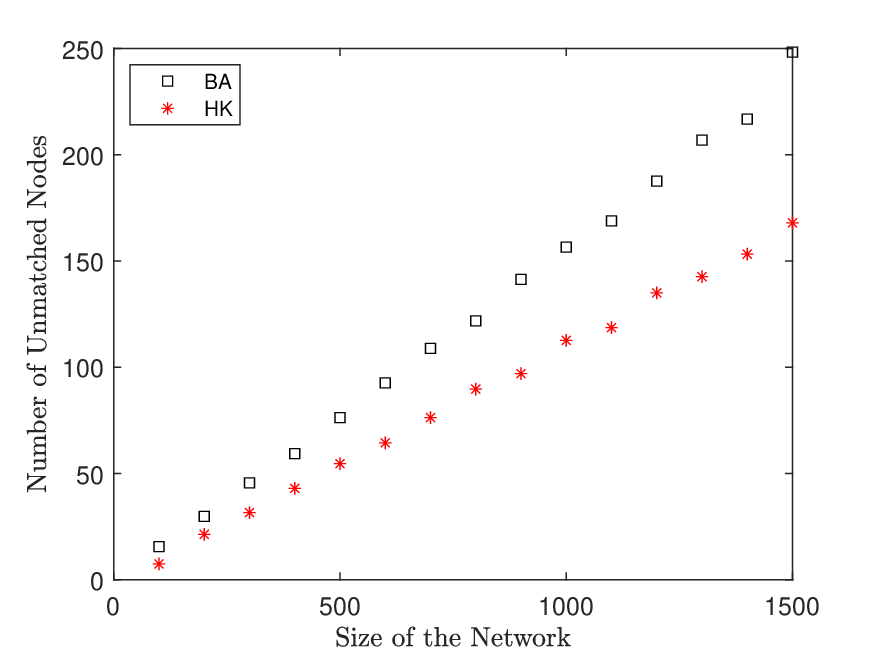}
	\caption{This figure compares the number of unmatched nodes of the BA and HK networks of the same size.}
	\label{fig_unmatched_size}
\end{figure}

\begin{figure}[hbpt!]
	\centering
	\includegraphics[width=3in]{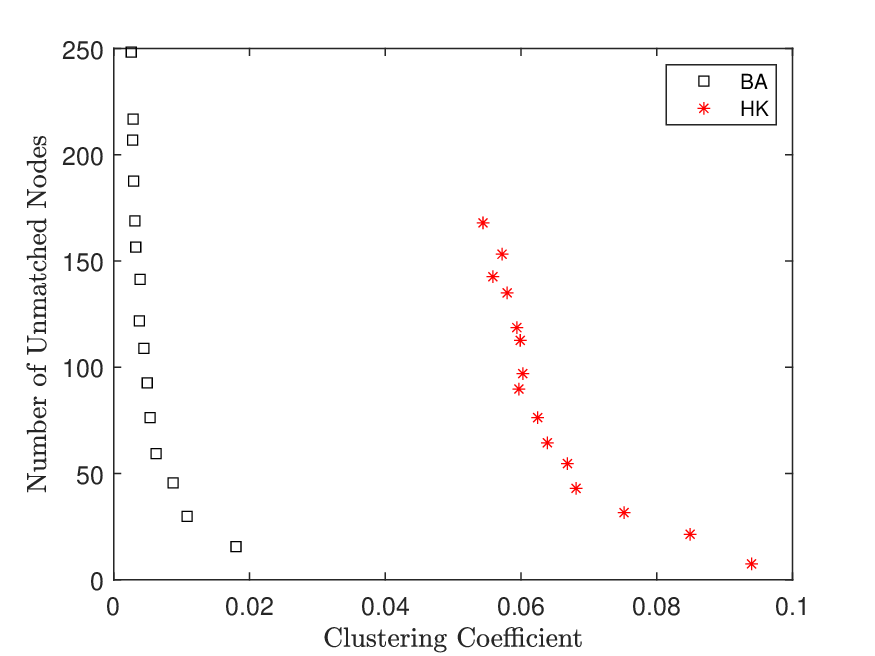}
	\caption{This figure compares the number of unmatched nodes versus the clustering coefficient of the BA and HK networks of the same size.}
	\label{fig_unmatched_clust}
\end{figure}

For simulation, we generate BA networks with $m=4$ links and similar HK models with $p=50\%$, i.e., $m=4$ and $m_1=2$. Note that, since $m=4$ for both models, the number of links and average node degree are the same (equal to $2m$ for $n \gg m$). We change the size of the network from $100$ nodes to $1500$ nodes and repeat this over $50$ iterations for Monte-Carlo simulation. We compare the properties of the two models, including the clustering coefficient in Fig.~\ref{fig_clust_size} and number of unmatched nodes in Fig.~\ref{fig_unmatched_size}.  
From Fig.~\ref{fig_clust_size} it is clear that the clustering coefficient of the HK model is more than the BA model of the same size; this is also shown theoretically via Eqs. \eqref{eq_C_BA} and \eqref{eq_C_HK}. The key result is represented in Fig.~\ref{fig_unmatched_size}. From the figure, it can be seen that  BA networks have more unmatched nodes than HK networks of the same size. Since other properties of the two networks (including the network size, number of links, and average node degree) are the same, this implies that there is an inverse correlation between the clustering coefficient and number of unmatched nodes in the network. In other words, increasing the clustering coefficient leads to decrease in unmatched nodes. This is also illustrated in Fig.~\ref{fig_unmatched_clust}.

One motivating outcome of these empirical results is that one can manage the controllability/observability of synthetic networks based on their number of unmatched nodes by tuning the clustering of those networks. For example, one may add few extra links to increase the triangle formations in the network to reduce its number of unmatched nodes. This is better illustrated in the next section. 

\section{Case Studies} \label{sec_case}
\begin{figure}[hbpt!]
	\centering
	\includegraphics[width=3in]{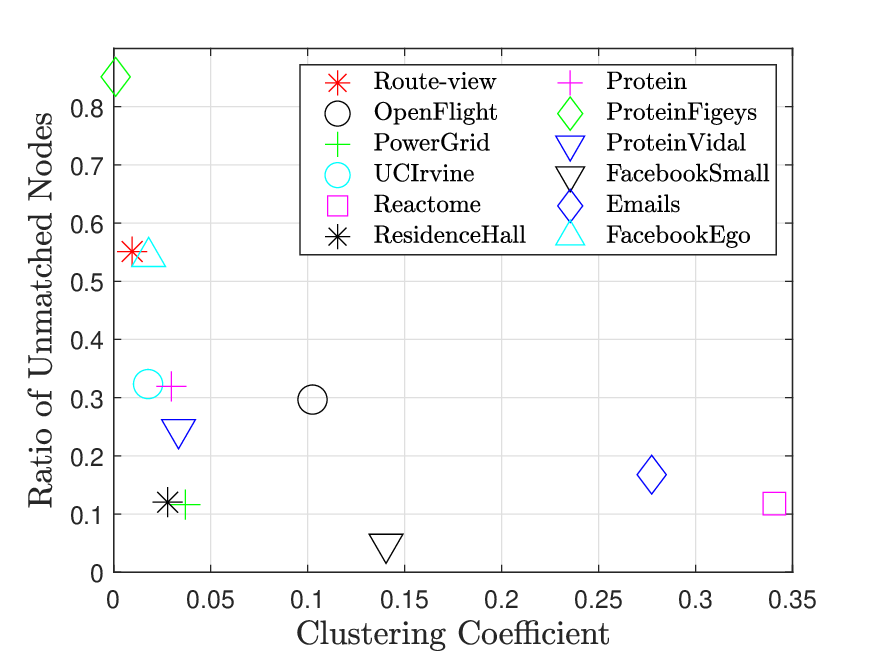}
	\caption{This figure presents the ratio of unmatched nodes to network size versus the clustering coefficient of some real-world network data.}
	\label{fig_netdata}
\end{figure}

In this section, we first compare different real-world networks in terms of clustering coefficient versus the ratio of unmatched nodes to the network size. The network data can be found in \cite{konect,UCI,ASU}. Fig.~\ref{fig_netdata} shows this comparison. 
As we observe from the figure, as a general trend, higher clustering coefficients correlate with lower ratios of unmatched nodes and vice-versa. This holds for many network cases in the figure. However, outliers or exceptions in specific networks exist. This is because different types of networks (e.g., social networks, biological networks, transportation networks) can have distinct structures and connectivity patterns and not necessarily follow SF structure and power-law degree distribution. The other factor is network connectivity (based on average node degree) which can significantly influence both the clustering coefficient and the number of unmatched nodes. Recall that in Section~\ref{sec_mc} for Monte-Carlo simulation both BA and HK networks follow SF structure with the same number of links and network average degree. However, this does not necessarily hold for real networks in this section. 

\begin{figure}[hbpt!]
	\centering
	\includegraphics[width=3in]{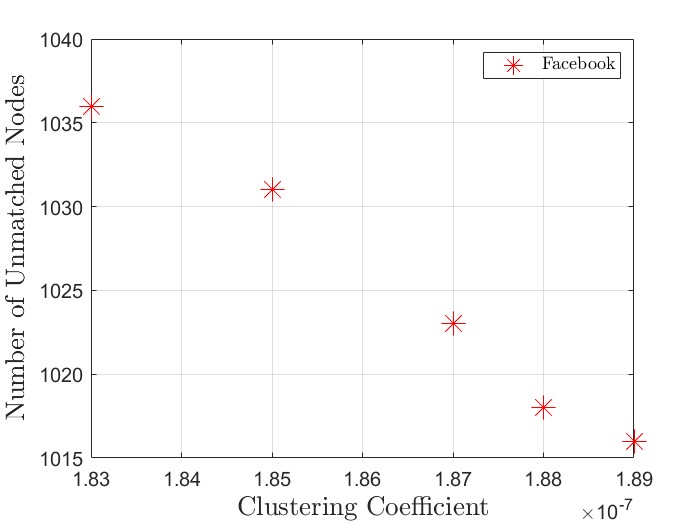}
	\caption{The change in the number of unmatched nodes by increasing the clustering coefficient via link addition in Facebook network.}
	\label{fig_facebook}
\end{figure}

Next, as a network tuning case study, we change the clustering coefficient of a real network by adding new links, and we check how the number of unmatched nodes changes. For this example, we consider the Facebook data set in \cite{konect} as a real social network. This network has $n=1911$ nodes and $5582$ links.
We use the following algorithm to iteratively increase its clustering coefficient by adding extra links:
\begin{enumerate}
	\item \textbf{Initialization:} Find the dilations $\mc{D}_i$ in the graph $\mc{G}$ via Dulmage-Mendelsohn decomposition.  
	\item \textbf{Link Addition:} Find two successive nodes $a,b$ in a dilation $\mc{D}_i$ and make a triangle by adding a link between these two nodes $a$ and $b$. 
	\item \textbf{Repeat:} Continue adding links until reaching the desired clustering or intended number of extra links.	
\end{enumerate}
This algorithm increases the clustering coefficient by making extra triangular motifs. 
The following table summarizes the outcome of this algorithm to increase the clustering coefficient and change in the number of unmatched nodes.
 \begin{table} [hbpt!]
	\centering
	\caption{Number of unmatched nodes versus clustering coefficient based on added number of extra links in Facebook network.}
	\begin{tabular}{|c|c|c|}
		\hline
		Added links& Unmatcheds & Clustering coefficient\\
		\hline
		$0$ & ~$1036$  &~$18.3 \times 10^{-3}$  \\
		\hline
		$5$ & ~$1031$  &~$18.5 \times 10^{-3}$  \\
		\hline
		$10$ & ~$1023$  &~$18.7 \times 10^{-3}$ \\
		\hline
		$15$ & ~$1018$  &~$18.8 \times 10^{-3}$ \\
		\hline
		$20$ & ~$1016$  &~$18.9 \times 10^{-3}$ \\
		\hline
		\hline
	\end{tabular}
	\label{tab_reduce}
\end{table} 
The results are also illustrated in Fig.~\ref{fig_facebook}.
Clearly from Table~\ref{tab_reduce}, by applying this algorithm, the clustering coefficient is increased and the number of unmatched nodes is decreased. Note that, in order to maintain the key network properties such as average node degree and degree distribution, we add few links which are ignorable as compared to the overall number of links in the network (implying almost the same average node degree).

\begin{figure}[hbpt!]
	\centering
	\includegraphics[width=3in]{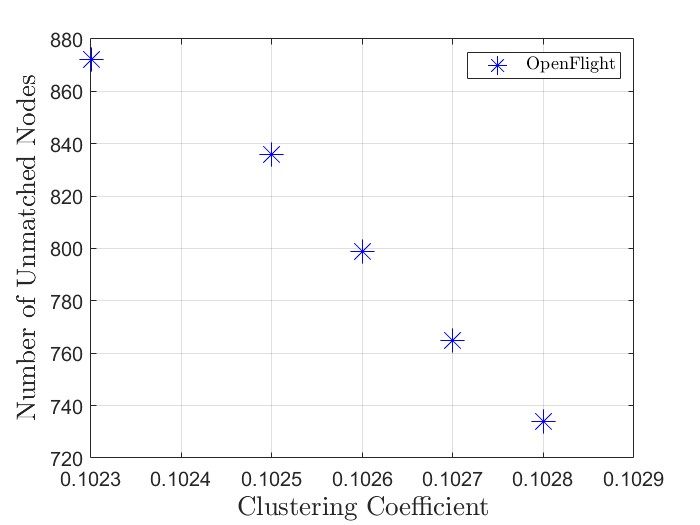}
	\caption{The change in the number of unmatched nodes by increasing the clustering coefficient via link addition in OpenFlight network.}
	\label{fig_openflight}
\end{figure}

As another example, we consider the OpenFlight transportation network in \cite{konect} with $n=2939$ nodes and $30501$ links. We increase the clustering coefficient of this network using the same dilation-based algorithm. The results are given in Table~\ref{tab_reduce2}.   
 \begin{table} [hbpt!]
	\centering
	\caption{Number of unmatched nodes versus clustering coefficient based on added number of extra links in OpenFlight network.}
	\begin{tabular}{|c|c|c|}
		\hline
		Added links& Unmatcheds & Clustering coefficient \\
		\hline
		$0$ & ~$872$  &~$102.3 \times 10^{-3}$  \\
		\hline
		$20$ & ~$836$  &~$102.5 \times 10^{-3}$  \\
		\hline
		$40$ & ~$799$  &~$102.6 \times 10^{-3}$ \\
		\hline
		$60$ & ~$765$  &~$102.7 \times 10^{-3}$ \\
		\hline
		$80$ & ~$734$  &~$102.8 \times 10^{-3}$ \\
		\hline
		\hline
	\end{tabular}
	\label{tab_reduce2}
\end{table}
The results are also illustrated in Fig.~\ref{fig_openflight}.
It is clear that by increasing the clustering coefficient, the number of unmatched nodes is decreased. This implies less number of observer/driver nodes for the observability/controllability of this network.

\section{Conclusion} \label{sec_con}
This study shows the key role of clustering in enhancing the observability and controllability of complex SF networks. By setting unmatched nodes in the network as the minimum number of required driver and observer nodes, our results show that highly-clustered SF networks require less number of driver/observer nodes. We should emphasize that the results in this work are stated for SF networks, and studying other types of network structures (e.g., small-world networks, Erdos-Renyi graphs, geometric graphs, and regular-lattices/spatial networks) are directions of future research.
	 
Based on our results one can reduce the number of unmatched nodes and the associated number of sensor/actuator nodes by increasing the clustering coefficient, for example, in transportation and industrial networks. This implies that one can facilitate network observability/controllability by tuning the clustering, thereby optimizing control/estimation strategies in many system applications. On the other hand, it reflects the implication of network topology on security, as evidenced by some literature \cite{zografopoulos2021cyber}.
	
\section*{Acknowledgements}
The authors acknowledge the use of some MATLAB codes from Koblenz Network Collection (KONECT) \cite{konect}. This work is funded by Semnan University, research grant No. 226/1403/1403208.
	
\section*{Statements \& Declarations}

This work is funded by Semnan University, research grant No. 226/1403/1403208.

The authors have no relevant financial or non-financial (or other competing) interests to disclose.

All authors contributed to the study conception, preparation, and writing.
All authors read and approved the final manuscript.

This research involves NO human or animal subjects.

Ethics declaration: not applicable

Data sharing not applicable to this article as no data-sets were generated or analysed during the current study.
	
\bibliographystyle{spmpsci} 
\bibliography{bibliography}
	
\end{document}